\documentclass[10pt,a4paper,twocolumn,aps,prb,nobibnotes,floatfix,superscriptaddress,showpacs]{revtex4-1}

\usepackage[utf8]{inputenc}
\usepackage[english]{babel}

\usepackage{amsfonts}
\usepackage{amsmath}
\usepackage{amssymb}
\usepackage{bbold}
\usepackage[dvipsnames,svgnames,table]{xcolor}
\usepackage{graphicx}
\usepackage{fancyhdr}
\usepackage{float}
\usepackage{braket}
\usepackage{float}
\usepackage[caption=false]{subfig}
\usepackage{tikz}  % drawings
\usepackage{chngcntr}

\usepackage{hyperref}
\hypersetup{
    pdfstartview={FitH},% fits the width of the page to the window
    colorlinks=true,    % false: boxed links; true: colored links
    linkcolor=NavyBlue, % color of internal links
    citecolor=Maroon,   % color of links to bibliography
    filecolor=NavyBlue, % color of file links
    urlcolor=NavyBlue   % color of external links
}

\makeatletter
\def\footnoterule{\kern -10pt
    \hrule \@width 100pt \kern 10pt} % the \hrule is .4pt high
\makeatother

\let \bm \boldsymbol
\def \newpar {\vspace{15pt}}
\def \angstrom {\text{\normalfont\AA}}
\def \dagg {^\dagger}
\def \ii {\mathfrak{i}}

\let \oldref \ref
\def \ref #1 {(\oldref{#1})}

\def \ufro {Departamento de Ciencias Físicas, Universidad de La Frontera, Casilla 54-D, 4811186 Temuco, Chile.}
\def \fcfm {Departamento de Física, FCFM, Universidad de Chile, Santiago, Chile.}
\def \aysen {Universidad de Aysén, Calle Obispo Vielmo 62, Coyhaique, Chile}
\def \cedenna {CEDENNA, Universidad de Santiago de Chile, Avda. Ecuador 3493, Santiago, Chile.}
\def \utrecht {Institute for Theoretical Physics, Utrecht University, 3584CC Utrecht, The Netherlands}
\def \norway {Center for Quantum Spintronics, Department of Physics,
Norwegian University of Science and Technology, NO-7491 Trondheim, Norway}
\def \eindhoven {Department of Applied Physics, Eindhoven University of Technology,\\
P.O. Box 513, 5600 MB Eindhoven, The Netherlands}

\begin{document}
    \title{Magnon Polarons induced by a magnetic field gradient}
    \author{N. Vidal-Silva}
        \email[Corresponding author: ]{nicolas.vidal@ufrontera.cl}
        \affiliation{\ufro}
        \affiliation{\cedenna}
    \author{E. Aguilera}
        \affiliation{\fcfm}
    \author{A. Roldán-Molina}
        \affiliation{\aysen}
    \author{R. A. Duine}
    \affiliation{\utrecht}
    \affiliation{\norway}
    \affiliation{\eindhoven}
    \author{A. S. Nunez}
     \affiliation{\cedenna}
    \affiliation{\fcfm}

    \begin{abstract}
        In this work, we report the theoretical possibility of generating magnon polaron excitations through a space-varying magnetic field.  The spatial dependence of the magnetic field in the Zeeman interaction gives rise to a magnon-phonon coupling when a magnetic field gradient is applied, and such a coupling depends directly on the strength of the gradient.  It is also predicted that the direction of the magnetic field gradient allows control over which phonon polarization couples to the magnons in the material.  Here we develop the calculations of the magnon-phonon coupling for an arbitrary (anti)ferromagnet, which are later used to numerically study its consequences.  These results are compared to the ones obtained with the phenomenological magnetoelastic coupling in YIG, where we show that the magnon polaron bandgap seen in YIG can be also obtained with a magnetic field gradient of $\sim 0.1$T/m which can be achieved with the current experimental techniques.  Our results propose a new way of controlling the magnetoelastic coupling in an arbitrary material and open a new route to exploit the magnon-phonon interaction in magnonic and spintronic devices.
    \end{abstract}

\maketitle

\section{Introduction}
In the last years, the magnetoelastic coupling has gained much attention due to the potential applications it offers in the field of spintronics \cite{vzutic2004spintronics,bader2010spintronics,chumak2015magnon}, magnonics \cite{kruglyak2010magnonics,serga2010yig}, spin caloritronics \cite{bauer2012spin} and more recently, spin-mechatronics \cite{matsuo2017spin}.  The simultaneous excitation of spin and elastic waves mediated by the magnetoelastic coupling gives rise to the so-called magnetoelastic waves \cite{akhiezer1959,scott1977}, which has been a focus of study over the past decades \cite{abrahams52,kittel53,kittel58}. However, due to recent progress in the synthesis and characterization techniques of materials, the effects related to the magnetoelastic coupling have been experimentally addressed only recently \cite{bozhko2020,park2020,zhang2019,berk2019}.

\newpar
From a quantum mechanical point of view, both spin and elastic waves have a quantized form of their elementary excitations, namely, magnons and phonons respectively.  Due to their bosonic nature, both quasiparticles obey the Bose-Einstein statistics.  In the long wave-length limit, magnons are usually charactacterized by their quadratic dispersion relation, which posseses a band-gap proportional to the external magnetic field and the magnetic anisotropy\cite{stancil2009}. On the other hand, phonons have a well-known linear dispersion at low energy and, according to the symmetry of the material, have three distinct vibrational modes \cite{srivastava2019}.  In absence of magnetoelastic coupling, the dispersion relations of magnon and phonon might cross at some wave-vector $\bm{k}^{*}$.  However, in the presence of magnetoelastic coupling, the interaction between magnons and phonons avoids the \textit{crossing point} at $\bm{k}^*$, and instead form what is called an \textit{anti-crossing point} \cite{gurevich1996,guerreiro2015,ruckriegel2014,flebus2017}.  At this point, the interaction between magnons and phonons is maximum and the related eigenstates are a hybridization between magnons and phonons, called magnon polarons or magnetoelastic waves\cite{flebus2017, kamra2015,shen2015}.

\newpar
Magnon polarons have recently been studied in the context of transport, topological and magnetic properties of, mainly, (anti)ferromagnetic insulators.  For example, anomalies in the Spin Seebeck \cite{kikkawa2016} and Spin Peltier \cite{yahiro2020} effect have been attributed to the presence of magnon polarons.  Local \cite{flebus2017} and non-local \cite{cornelissen2017,rameshti2019} magnon polaron spin transport has been also measured in YIG films.  More recently, the topological nature of magnon polarons has been predicted \cite{go2019,zhang20193,park2020,thingstad2019}, as well as the control of its topology \cite{shen2020, go2019}. Antiferromagnets also present magnon polarons, as reported in references \cite{simensen2019,sukhanov2019}.  Particularly, non-collinearity in antiferromagnets has been pointed out as a source of magnon polaron excitations \cite{sukhanov2019}.  Spin pumping has been also enhanced due to the presence of magnon polarons \cite{hayashi2018}. Thus, in most of the effects attributed to magnon polarons, the magnitude of its contribution depends essentially on the magnetoelastic parameter, which quantifies the strength of the interaction. For instance, in reference \onlinecite{nomura2019}, the non-reciprocity of the sound velocity in the Phonon Magnetochiral effect is mediated by the cubic of a magnetoelastic constant. In the same way, the magnon lifetime due to the phonon scattering is also proportional to the magnetoelastic constant \cite{ruckriegel2014,streib2019}. The enhancement of magnetization damping by phonon pumping has been reported to be proportional to the magnetoelastic constant too \cite{streib2018}. In general, any physical quantity related to the action of magnon polarons depends on a magnetoelastic parameter. Importantly, the strength and source of this interaction are not unique: while there is an intrinsic anisotropy-mediated magnetoelastic coupling, hereafter phenomenological magnetoelastic coupling, which stems from the spin-orbit coupling and dipole-dipole interaction \cite{kittel53,kittel58}, there are also another sources of magnetoelastic coupling as the dependence of the exchange interaction on the lattice deformations \cite{gurevich1996,streib2019,ruckriegel2014,maehrlein2018} or the modulation of the Dzyaloshinskii-Moriya interaction by shear strains \cite{nomura2019,zhang2019,park2020}.

\newpar
In this work, we study how a magnetoelastic coupling can be induced by applying a magnetic field gradient on a arbitrary magnetic lattice. We will show analytically and numerically that the coupling depends directly on the magnitude and direction of the magnetic field gradient. This will be shown to imply that the experimental control of the magnetic field's shape allows the tuning of the coupling strength and the possibility of selecting which phonon polarization couples to the magnons of the material.  The presented coupling could be applied, in principle, to any (anti)ferromagnetic lattice with a crossing point between the dispersion relations of magnons and phonons.

\newpar
The paper is organized as follows: in section \ref{model} we start our study by describing the proposal with a toy model in a uni-dimensional system. Despite this is a pretty simple model, it will allow us to establish the role that a magnetic field gradient plays in the stability of a given system in the presence of an inhomogeneous magnetic field with the same periodicity of the lattice. Once we identify the conditions that our system must have in order to be physically realizable, in sections \ref{magnons} and \ref{phonons} we introduce the basic concepts on the quantization of the magnetic and elastic systems in a superlattice, respectively, and also explore the analytical nature of the magnon-phonon coupling due to a magnetic field gradient in section \ref{magnon_polarons}. Also, in section \ref{numerical_calculations}, we detail the numerical algorithm we used to diagonalize the Hamiltonian of our system. Next, in section \ref{MPbands} we apply our results to a Magnonic Crystal, where we study the dispersion relation of magnon polarons and highlight the main properties of the energy bands obtained with the proposed coupling mechanism.  We also make a comparison between the phenomenological magnetoelastic coupling and the one we propose.  Finally, in section \ref{discussion}, we discuss and give some conclusions for future works.

\section{Model}
\label{model}

In this section, we will describe the nature of magnons and phonons in an arbitrary lattice, as well as their coupling due to the enforcement of a magnetic field gradient in the presence of a space-varying magnetic field.  For simplicity, we will assume a low-temperature regime such that the magnetization's fluctuations are weak enough to keep the magnetic order with no thermal disturbance.  The idea of this section is to capture the physics behind the magnetic field gradient-mediated magnon-phonon coupling considering an inhomogeneous magnetic field with the same periodicity of the lattice.  This will allow us to understand the physical limitations of the proposal, and it will also pave our way to the next section, where we will overcome some of these limitations by changing the spatial-periodicity of the magnetic field.  We will also consider a ferromagnetic insulator to neglect the electronic charge.

\begin{figure}
    \includegraphics[width=0.43\textwidth]{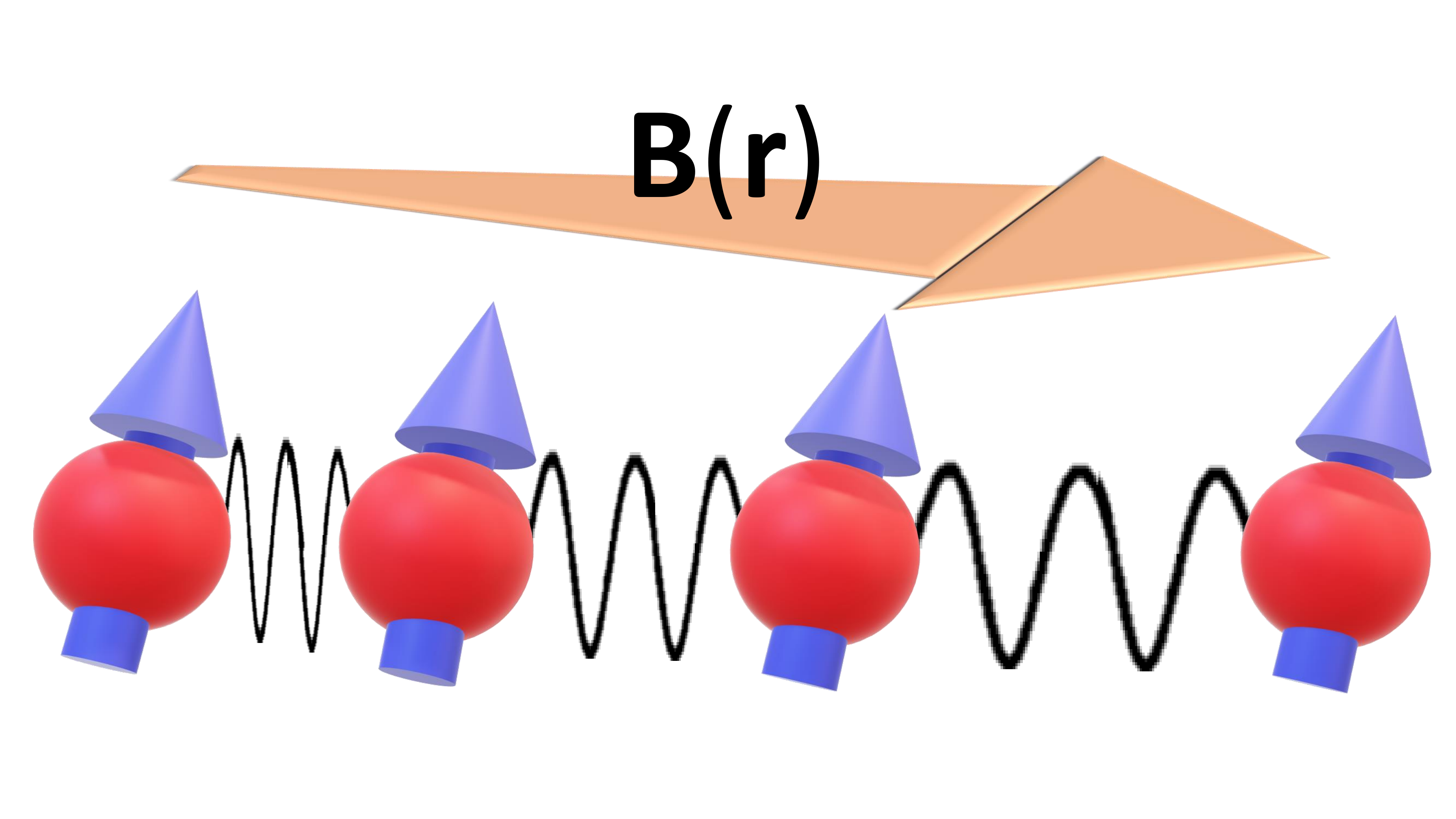}%
    \caption{Schematic representation of the unidimensional lattice with a space-varying magnetic field. The nonuniform arrow represents a magnetic field gradient, which according to our proposal, exerts an external force on each magnetic dipole which deviates them from its equilibrium position so ultimately excites simultaneously magnon and phonon modes, generating thus magnon polaron excitations.}
    \label{fig:drawing}
\end{figure}

\newpar
As mentioned before, we will begin by analysing a unidimensional magnetoelastic lattice with nearest neighbor distance $a_0$ in a space-varying magnetic field $\bm{B}(\bm{r})$ with the same periodicity of the lattice. For this first example, we will consider that the system is dominated by a nearest neighbor elastic coupling, nearest-neighbor Heisenberg exchange and Zeeman interaction, as depicted in Fig \ref{fig:drawing}. In this way, we consider a spin chain along the $x$ axis and parameterize the Hamiltonian of it in terms of the displacement $u_i = x_i - X_i$, being $x_i$ and $X_i$ the position and equilibrium position of the site $i$, respectively; the phonon momentum $p_i$, and the spin vectors $\bm{S}$, meaning that the Hamiltonian reads
\begin{align}
    \label{H1}
    \nonumber
    \mathcal{H} &= \sum_{i=1}^N \Bigg[
        \frac{{p_i}^2}{2M} + \frac{M{\omega_0}^2}{2} \big(u_{i+1} - u_i\big)^2
        \\ &
        - J \bm{S}_i \cdot \bm{S}_{i+1}
        - \mu_B g  \bm{B}(x_i)\cdot\bm{S}_i
    \Bigg]
    \text{ ,}
\end{align}
where $J$ is the Heisenberg exchange constant, $M$ is the average mass of each site, $\omega_0$ is the natural frequency of the elastic coupling between two neighboring sites, $\mu_B$ the Bohr magneton and $g$ the Landé factor.

\newpar
We can expand the magnetic field around its equilibrium $X_i$ up to to first order in the displacement $u_i$, in the displaced position $x_i=X_i+u_i$:
\begin{align}
    \label{B1_taylor}
    B^\alpha(x_i) =
        B^\alpha(X_i)
        + \frac{\partial B^{\alpha}}{\partial x} \Bigg|_{x=X_i} u_i + \mathcal{O}(u_i^2)
    \text{ .}
\end{align}
It must be noticed that we have assumed that the magnetic field gradient is weak enough to do not affect the equilibrium position, such that $X_i$ is independent of the gradient. Note that a magnetoelastic coupling has been induced, as evidenced in the linear term of the expansion in displacement $u_i$. This procedure will be used from now on and it will be the base to show how a magnetic field gradient drives an induced magnetoelastic coupling. Eq. \ref{B1_taylor} can be directly replaced on Hamiltonian \ref{H1} to obtain
\begin{align}
    \label{H1_taylor}
    \nonumber
    \mathcal{H} &= \sum_{i=1}^N \Bigg[
        \frac{{p_i}^2}{2M} + \frac{M{\omega_0}^2}{2} \big(u_{i+1} - u_i\big)^2
        - J \bm{S}_i \cdot \bm{S}_{i+1}
        \\ &
        - \mu_B g  \bm{B}(X_{i})\cdot\bm{S}_i
        - \mu_B g  u_i\bm{S}_i\cdot\frac{\partial \bm{B}}{\partial x}\Bigg|_{X_{i}}
    \Bigg]
    \text{ ,}
\end{align}
where the induced magnetoelastic interaction becomes evident in the last term of the above expression when coupling the elastic degrees of freedom $\{
u_i\}$ with the magnetic ones $\{\bm{S}_i\}$.
\newpar

Before proceeding onto studying the coupling of magnons and phonons, it is essential to analyze the classical equilibrium of the system. This study is crucial because a magnetic field gradient exerts a force on every magnetic dipole, which could change the behavior of its equilibria. To study the equilibria of Hamiltonian \ref{H1}, the spin variable will be written in term of
its spherical angles $\phi_i$ and $\theta_i$ and it will be assumed that the magnetic field is given by $\bm{B}(x) = (B^x(x),0,0)$. The main idea is then to write down the Hamiltonian \ref{H1} as a function of the variables $\{\theta_i,\phi_i, u_i\}$ and after Fourier transform the resulting Hamiltonian, minimize it respect to the variables $\{u_k,u_{-k},\theta_k,\theta_{-k}\}$. Major details about the equilibrium analysis at this stage can be found in the Appendix \ref{appendix_equilibrium}. Thus, it can be proven that every eigenvalue of Eq. \ref{H1} is positive if and only if

\begin{align}
    \label{equilibrium_condition}
    \nonumber
    \Bigg[S g \mu_B \frac{\partial B^x}{\partial x}\Bigg]^2 &<
        4 S M{\omega_0}^2\Bigg[-2 J S
    \\  & \hspace{-30pt}
        + 2 J S\cos(a_0k)\Big)
        + \mu_B g B_z\Bigg]\sin^2\Bigg(\frac{ka_0}{2}\Bigg)
    \text{ .}
\end{align}

\newpar
It is essential to recall that an stable equilibrium is obtained if every eigenvalue of the Hessian is positive.  In this particular case, it must be noted that there exists a value of $k$ such that inequality \ref{equilibrium_condition} is not satisfied, meaning that the system is not in the real ground state, and then it is unstable. This can be understood in terms of a net force felt by \textit{each site} in the lattice. In this sense, a magnetic field gradient acts as an external force which ultimately accelerates the system. An accelerated system is no longer at stable equilibrium and then the real equilibrium state must be achieved. Eq. \ref{equilibrium_condition} says that for some range of $k$ values this new equilibrium state will never be reached, which means that magnons and phonons at that regime are unstable and no magnetoelastic coupling it would be observed, as shown in Fig. \ref{Appendix_fig1} of the appendix \ref{appendix_equilibrium}. Thus, by choosing a magnetic field with the same periodicity of the lattice, whatever be the magnetic field gradient applied, the spins equilibrium position or the magnetic ground state, the system will always show nonequilibrium aspects and then magnon polaron excitations are not allowed.
\newpar

To overcome this issue and obtain stable magnon polarons in the entire first Brillouin zone, we are going to study the problem of these hybridized quasiparticles in an arbitrary lattice composed of $N$ unit cell with a basis of $m$ sites, which will allow us to adapt the magnetic field to get a stable configuration. In other words, we will adjust the magnetic field periodicity such that the net force, that emerges from the gradient, acts now on \textit{each cell} of the system and to be identically nulled.

\newpar
We will separate the study of the total Hamiltonian in a arbitrary lattice into three partial Hamiltonians:
\begin{equation}
    \mathcal{H} = \mathcal{H}_m + \mathcal{H}_{ph} + \mathcal{H}_{mp}
    \text{ .}
\end{equation}
In Hamiltonian $\mathcal{H}_m$ we will include the magnon terms, which will come from the Heisenberg exchange and the zeroth order expansion of the magnetic field in the Zeeman term.  In $\mathcal{H}_{ph}$ we will consider the purely phononic terms, which come from the kinectic energy and a elastic potential.  $\mathcal{H}_{mp}$ includes the term that couple magnons and phonons, which will come from the first order expansion in the displacement of the Zeeman energy.

\subsection{Magnons}
\label{magnons}

Magnons are the bosonic elementary excitations of magnetic order and they are usually interpreted as the quanta of spin waves\cite{holstein40}. This system will be under the influence of an anisotropic exchange interaction and the Zeeman interaction with the external magnetic field.  With this, we have that the Hamiltonian reads
\begin{align}
    \label{H_magnetic}
    \mathcal{H} &= -\frac{1}{2}\sum_{ii'jj'}  S_{ij}^{\alpha} J_{i-i'}^{jj' \alpha \beta} S_{i'j'}^{\beta}
        -\mu_B g \sum_{ij} \bm{B}(\bm{r}_{ij}) \cdot \bm{S}_{ij}
    \text{ ,}
\end{align}
where summation over repeated greek indices is implied throughout this article and in this case  $\alpha,\beta \in \{\bm{\hat{x}}, \bm{\hat{y}}, \bm{\hat{z}}\}$.  The indices $i,i' \in \{1,2,...,N\}$ and $j,j' \in \{1,2,...,m\}$ represent the unit cells and basis sites respectively, and $i-i' \equiv \bm{R}_i - \bm{R}_{i'}$ is the distance between nearest neighbors unit cells. Note that the quantities with subindices $ij$ should be understood as the $jth$ element (basis site) of the $ith$ unit cell of the system. In the Hamiltonian we have also included the tensor $J_{i-i'}^{jj'}$, which corresponds to a generalized interaction between sites $\bm{S}_{ij}$ and $\bm{S}_{i'j'}$ with no particular choice of a given symmetry such that it might contain as the nearest neighbors exchange interaction as well as a Dzyaloshinskii-Moriya interaction.

\newpar
In order to isolate the terms purely related with the magnetic degrees of freedom from the Zeeman term, we will proceed as in Eq. \ref{B1_taylor} and expand the magnetic field around the equilibrium positions $\bm{R}_i$ upto first order in their displacement $\bm{u}_i$ as
\begin{equation}
    \label{B_taylor}
    B^\alpha(\bm{r}_{ij}) =
        B^\alpha(\bm{R}_{ij})
        + \frac{\partial B^{\alpha}}{\partial r^{\beta}} \Bigg|_{\bm{r}_{ij}=\bm{R}_{ij}}u_{ij}^{\beta} + \mathcal{O}({\bm{u}_{ij}}^2)
    \text{ ,}
\end{equation}
where we are going to keep only the first term of the expansion and in section \ref{magnon_polarons} we are going to consider the second one to obtain the magnon-phonon coupling.  We will also adopt the notation
\begin{align}
    \label{B_notation}
    B_j^\alpha &\equiv B^\alpha(\bm{R}_{ij})
    &\text{ and} &  &
    {B'}_j^{\alpha\beta} &\equiv \frac{\partial B^{\alpha}}{\partial r^{\beta}} \Bigg|_{\bm{r}_{ij}=\bm{R}_{ij}}
    \text{ .}
\end{align}

\newpar
To obtain a quantized magnonic Hamiltonian we must start by using the Holstein-Primakoff transformation \cite{holstein40}, which allow us to write the the spin operators $\bm{S}_{ij}$ in term of bosonic operators $a_{ij}$ and $a\dagg_{ij}$, which annihilates and creates magnons, respectively.  This transformation reads
\begin{subequations}
    \label{holstein_primakoff}
    \begin{align}
        S_{ij}^x &\approx \sqrt{\frac{S}{2}} \Big( a\dagg_{ij} + a_{ij} \Big)
        \\
        S_{ij}^y &\approx \ii \sqrt{\frac{S}{2}} \Big( a\dagg_{ij} - a_{ij} \Big)
        \\
        S_{ij}^z &= S - a\dagg_{ij} a_{ij}
        \text{ ,}
    \end{align}
\end{subequations}
where we have already expanded upto second order in the magnon operators as they are the only terms we will be dealing with in this article.  Furthermore, we are interested in obtaining the description of magnons in $\bm{k}$-space, which is obtained by employing the Fourier series, given by
    \begin{align}    \label{magnon_bloch}
        a_{ij} &= \frac{1}{\sqrt{N}}\sum_k a_{kj} e^{\ii \bm{k} \cdot \bm{r}_{ij}}
        \text{.}
    \end{align}
Now, we can simultaneously replace equations \ref{holstein_primakoff} and \ref{magnon_bloch} into Hamiltonian \ref{H_magnetic} to obtain the Hamiltonian for magnons in $\bm{k}$-space, which reads
\begin{widetext}
    \begin{align}
        \label{H_m}  \nonumber
        \mathcal{H}_m &= - \frac{S}{4} \sum_{jj'\bm{k}} \Bigg[
            \Gamma_{\bm{k}}^{jj'-} a\dagg_{\bm{k}j}a\dagg_{-\bm{k}j'}
            + \bar{\Gamma}_{\bm{k}}^{jj'-} a_{-\bm{k}j}a_{\bm{k}j'}
        % \\ \nonumber &
            + \Gamma_{\bm{k}}^{jj'+}  a\dagg_{\bm{k}j}a_{\bm{k}j'}
            + \bar{\Gamma}_{\bm{k}}^{jj'+} a_{-\bm{k}j}a\dagg_{-\bm{k}j'}
        % \\ \nonumber &
            -2 J_{0}^{jj'zz} \Big( a\dagg_{\bm{k}j}a_{\bm{k}j} + a\dagg_{\bm{k}j'}a_{\bm{k}j'} \Big)
            \Bigg]
        \\ &
        + \mu_B g \sum_{j\bm{k}} B_j^z a\dagg_{\bm{k}j}a_{\bm{k}j}
        \text{ ,}
    \end{align}
\end{widetext}
where $\Gamma_{\bm{k}}^{jj'\pm}$ and $\bar{\Gamma}_{\bm{k}}^{jj'\pm}$ are defined as
\begin{subequations}
    \label{def_Gamma}
    \begin{align}
        \Gamma_{\bm{k}}^{jj'\pm} &=
            J_{\bm{k}}^{jj'xx} \mp \ii J_{\bm{k}}^{jj'xy} + \ii J_{\bm{k}}^{jj'yx} \pm J_{\bm{k}}^{jj'yy}
        \text{ ,}  \\
        \bar{\Gamma}_{\bm{k}}^{jj'\pm} &=
            J_{\bm{k}}^{jj'xx} \pm \ii J_{\bm{k}}^{jj'xy} - \ii J_{\bm{k}}^{jj'yx} \pm J_{\bm{k}}^{jj'yy}
        \text{ ,}
    \end{align}
\end{subequations}
and
\begin{align}
    \label{fourier_J}
    J_{\bm{k}}^{jj'\alpha\beta} &= \sum_k J_{i-i'}^{jj'\alpha\beta} e^{\ii \bm{k} \cdot ( \bm{r}_{ij}-\bm{r}_{i'j'} )}
    \text{ .}
\end{align}

\newpar
The result obtained in equation \ref{H_m} can be used for any lattice with magnetic order.  Eventhough, in equation \ref{holstein_primakoff} we have assumed that the magnetic order in equilibrium is equal to $\bm{\hat{z}}$ for every site in the lattice, we can incorporate any periodic magnetic texture described by the equilibriums $\bm{S}_{0}(\theta_j, \phi_j)$ (for instance a skyrmion or vortex lattice) by introducing a local change of coordinates at every site by means of rotation matrices
\begin{equation}
    \label{rotation_matrix}
    R_j \equiv \begin{pmatrix}
            \cos \phi_j & -\sin \phi_j & 0 \\
            \sin \phi_j &  \cos \phi_j & 0 \\
                      0 &           0 & 1 \\
        \end{pmatrix} \begin{pmatrix}
             \cos \theta_j & 0 &  \sin \theta_j \\
                         0 & 1 &              0 \\
            -\sin \theta_j & 0 &  \cos \theta_j \\
        \end{pmatrix}
    \text{,}
\end{equation}
which can be used to redefine the anisotropic exchange tensor as
\begin{equation}
    \label{J_rotation}
    \mathcal{J}_{i-i'}^{jj'} = {R_j}^T  J_{i-i'}^{jj'} R_{j'}
    \text{ ,}
\end{equation}
where we only have to use $\mathcal{J}_{i-i'}^{jj'}$ instead of $J_{i-i'}^{jj'}$ in equation \ref{fourier_J}.

\subsection{Phonons}
\label{phonons}

Analog to spin waves, elastic waves can also be quantized. The elementary excitations of elastic waves are the so-called phonons. To describe phonons in our system, we consider that each ion with mass $M$ at position $\bm{r}_{ij}$ deviates from its equilibrium position $\bm{R}_{ij}$ by a small displacement $\bm{u}_{ij} = \bm{r}_{ij} - \bm{R}_{ij}$, such that the phonon Hamiltonian can be written as \cite{blundell2003,marder2010}
\begin{equation}
    \label{H_elastic}
    \mathcal{H}_{ph} = \sum_{ij} \frac{{\bm{p}_{ij}}^2}{2M}
    + \sum_{ii'jj'} \frac{1}{2} u_{ij}^\alpha \Phi_{i-i'}^{jj'\alpha\beta} u_{i'j'}^\beta
    \text{ ,}
\end{equation}
where $\bm{p}_i$ is the conjugate momentum vector and the index convention is the same as used for equation~\ref{H_magnetic}.  Additionally we have that the mechanical interaction between two sites is described by the elastic tensor $\Phi_{i-i'}^{jj'\alpha\beta}$, which can be used to define
\begin{align}
    \label{fourier_Phi}
    \Phi_{\bm{k}}^{jj'\alpha\beta} &= \sum_{\bm{k}} \Phi_{i-i'}^{jj'\alpha\beta} e^{\ii \bm{k} \cdot ( \bm{r}_{ij}-\bm{r}_{i'j'} )}
    \text{ .}
\end{align}
It is important to note that because $\Phi_{i-i'}^{jj'\alpha\beta}$ is originally obtained from the second-order expansion of the potential energy between sites $\bm{r}_{ij}$ and $\bm{r}_{i'j'}$ \cite{marder2010}, we must have that $\Phi_{\bm{k}}^{jj'\alpha\beta}$ is real and symmetric, which ultimately implies that it is diagonalizable as
\begin{align}
    \Phi_{\bm{k}}^{\mu\nu} \epsilon_{\bm{k}\lambda}^\nu = \phi_{\bm{k}\lambda} \epsilon_{\bm{k}\lambda}^\mu
    \text{ ,}
\end{align}
where we have used the indices $\mu$ and $\nu$ as a short-hand to represent the basis $j, j'$ and coordinate $\alpha, \beta$ as a single index.  With the diagonalization of the problem, we obtain $\lambda \in \{1,2,...,3m\}$ eigenvalues and eigenvectors. Here, the vectors $\bm{\epsilon}_{\lambda}$ encode the phonon polarizations which can in turn be used to write the operators in $\bm{k}$-space using the discrete Fourier transform:
\begin{subequations}
    \begin{align}
    \label{fourier_up}
        u_i^\mu &= \frac{1}{\sqrt{N}} \sum_{\bm{k}\lambda} u_{\bm{k}\lambda} \epsilon_{\bm{k}\lambda}^\mu e^{\ii \bm{k} \cdot \bm{r}_{ij}}
        \text{ ,}  \\
        \label{fourier_pp}
        p_i^\mu &= \frac{1}{\sqrt{N}} \sum_{\bm{k}\lambda} p_{\bm{k}\lambda} \epsilon_{\bm{k}\lambda}^\mu e^{\ii \bm{k} \cdot \bm{r}_{ij}}
        \text{ .}
    \end{align}
\end{subequations}

Replacing equations \ref{fourier_up} into equation \ref{H_elastic} to obtain the Hamiltonian in $\bm{k}$-space yields
\begin{align}
    \label{Hel_k}
    H_{ph} &= \sum_{\bm{k}\lambda}  \Bigg[
        \frac{p_{\bm{k}\lambda}p_{-\bm{k}\lambda}}{2M}
        + \frac{M}{2}\omega_\lambda(\bm{k})u_{\bm{k}\lambda}u_{-\bm{k}\lambda}\Bigg]
        \text{ .}
\end{align}

To transform the displacement and momentum operators of Hamiltonian \ref{Hel_k} to phonon creation and annihilation operators, we will use the usual transformation
\begin{subequations}
    \label{phonons_operators}
    \begin{align}
        \label{uk_c}
        u_{\bm{k}\lambda} &= \sqrt{\frac{\hbar}{2M\omega_\lambda(\bm{k})}}
            \Big( c\dagg_{-\bm{k}\lambda} + c_{\bm{k}\lambda} \Big)
        \text{ ,}  \\
        \label{kk_c}
        p_{\bm{k}\lambda} &= i\sqrt{\frac{\hbar M\omega_\lambda(\bm{k})}{2}}
            \Big( c\dagg_{-\bm{k}\lambda} - c_{\bm{k}\lambda} \Big)
        \text{ ,}
    \end{align}
\end{subequations}
where the operator $c_{\bm{k}\lambda}\dagg$ ($c_{\bm{k}\lambda}$) creates (annihilates) a phonon with momentum $\bm{k}$ and polarization $\lambda$, and obey the commutation usual commutation relations
\begin{subequations}
    \begin{align}
        \big[c_{\bm{k}\lambda},c\dagg_{\bm{k}'\lambda'}\big] &= \delta_{\bm{k}\bm{k}'}\delta_{\lambda\lambda'}
        \text{ ,}  \\
        \big[c_{\bm{k}\lambda},c_{\bm{k}'\lambda'}\big] &= \big[c\dagg_{\bm{k}\lambda},c\dagg_{\bm{k}'\lambda'}\big] = 0
        \text{ .}
    \end{align}
\end{subequations}

Replacing the transformations \ref{phonons_operators} into Hamiltonian \ref{Hel_k}, we obtain an already diagonalized form of the phonon Hamiltonian.
\begin{align}
    \label{Hph_k}
    H_{ph} &= \sum_{\bm{k}\lambda}\hbar\omega_\lambda(\bm{k})  \bigg(c\dagg_{\bm{k}\lambda}c_{\bm{k}\lambda} + \frac{1}{2}\bigg)
    \text{ ,}
\end{align}
where the phonon's dispersion relation is
\begin{equation}
    \omega_\lambda(\bm{k}) = \sqrt{\frac{\phi_{\bm{k}\lambda}}{M}}% 2\hbar\sqrt{\frac{\phi_{\lambda}}{M}}  \Bigg| \sin\left(\frac{k a_0}{2}\right) \Bigg|%
   \text{,}
\end{equation}
being $\phi_{\bm{k}\lambda}$ the eigenvalues of the tensor $\Phi_{\bm{k}}^{jj'\alpha\beta}$. For simplicity, in the numerical calculations we show in the next section we will consider an isotropic material. Specifically, we will use sound velocities $v_{\lambda}$ reported for YIG samples, which are incorporated to the elastic tensor by setting its Fourier transformed equal to

\begin{eqnarray}
    \nonumber\Phi_{\bm{k}}^{jj'\alpha\beta} = V^{\alpha\beta}\Big(2-\left(1+e^{-\ii 2ka_0}\right)\delta_{j,0}\delta_{j',1}\\
    -\left(1+e^{\ii 2ka_0}\right)\delta_{j,1}\delta_{j',0}\Big)\text{ ,}
\end{eqnarray}
where 
\begin{equation}
    V^{\alpha\beta} = \frac{M}{{a_0}^2} \begin{pmatrix}
            {v_\parallel}^2 &           0 &               0  \\
                          0 & {v_\perp}^2 &               0  \\
                          0 &           0 & {v_\perp}^2
        \end{pmatrix}
    \text{ .}
\end{equation}
It is crucial to note this particular elastic tensor allows us to recover the well-known phonon's dispersion relation in the long wave-length limit, which is given by
\begin{equation}
    \hbar\omega_\lambda(\bm{k}) = v_{\lambda} |\bm{k}|
    \text{ ,}
\end{equation}
where $v_\parallel$ corresponds to the sound velocity of the longitudinal mode, while $v_\perp$ to the transversal mode.

\subsection{Magnon polarons}
\label{magnon_polarons}

The hybridization between magnons and phonons mediated by the magnetoelastic coupling forms the so-called magnon polarons. They are the quanta of the magnetoelastic waves which are a solution of the coupled set of differential equations involving the magnetic and elastic degrees of freedom \cite{gurevich1996}. The way we choose to obtain the magnon polaron excitations is to quantize the total Hamiltonian composed by the magnetic, elastic and magnetoelastic parts:
\begin{equation}
    \label{H_total}
    \mathcal{H} = \mathcal{H}_m + \mathcal{H}_{ph} + \mathcal{H}_{mp}
    \text{ .}
\end{equation}
where $\mathcal{H}_m$ is given by equation \ref{H_m} and $\mathcal{H}_{ph}$ by equation \ref{Hph_k}.  To obtain the expression of the magnon-phonon Hamiltonian $\mathcal{H}_{mp}$ we must recall that in the series expansion of the magnetic field in equation \ref{B_taylor}, the second linear term in the displacement was kept apart and it is the only term needed to obtain the magnon-phonon coupling.  Thus, the remaining Hamiltonian is
\begin{equation}
    \label{H_gradient}
    \mathcal{H}_{mp}    = - \mu_B g \sum_{ij} {B'}_j^{\alpha\beta} S_{ij}^\alpha u_{ij}^\beta
    \text{ .}
\end{equation}
where the derivative is evaluated in the equilibrium position $\bm{R}_{j}$ as established in Eq. \ref{B_notation}.

\newpar
To obtain the quantized form of the magnon-phonon Hamiltonian explicitly in the $\bm{k}$ space, we must start by using the Fourier transformation of $u_i$ given in equation \ref{fourier_up}.  Following this, we make use of the transformations given in equations \ref{holstein_primakoff} and \ref{uk_c} and the Bloch's theorem over the magnonic creation and annihilation operators to obtain the magnetoelastic Hamiltonian in second quantization, which reads
\begin{align}
    \label{Hmp_k}
    \mathcal{H}_{mp} &= \sum_{\bm{k}\lambda j}
        \bigg[\Lambda_{\bm{k}\lambda}a_{-\bm{k}j}\Big(c_{\bm{k}\lambda} + c\dagg_{-\bm{k}\lambda} \Big) + \text{h.c.}\bigg]
        \text{ ,}
\end{align}
with the interaction parameter $\Lambda_{\bm{k}\lambda}$ is
\begin{align}
    \label{lambdamec}
    \Lambda_{\bm{k}\lambda} &= -\mu_B g \sqrt{\frac{\hbar S}{4M\omega_\lambda(\bm{k})}}
        \Bigg(B'^{x\beta}_j- \ii B'^{y\beta}_j \Bigg) \epsilon_{\bm{k}\lambda}^\beta
    \text{ .}
\end{align}
From equations \ref{Hmp_k} and \ref{lambdamec} we can effectively see how a magnon-phonon coupling emerges and that this is proportional to the magnetic field gradient. Furthermore, the magnetoelastic parameter $\Lambda_{\bm{k}\lambda}$ depends essentially on the magnitude of the derivatives of the transverse components of the magnetic field.  More importantly, the gradient direction couples differently with each phonon polarization, which in this case correspond to the $x$, $y$ and $z$ axis.  This last point means that in principle, there is complete freedom to choose which phonon and magnon bands hybridize.  Comparing with the usual phenomenological magnetoelastic Hamiltonian\cite{flebus2017,streib2019,ruckriegel2014} given by
\begin{equation}
    \label{Hmp_k_2}
    \mathcal{H}_{mp}^{K} = \sum_{\bm{k}\lambda}
        \bigg[\Gamma_{\bm{k}\lambda} a_{-\bm{k}}\left(c_{\bm{k}\lambda} + c\dagg_{-\bm{k}\lambda}\right) + \text{h.c.}\bigg]
    \text{ ,}
\end{equation}
where
\begin{equation}
    \label{gammamec}
    \Gamma_{\bm{k}\lambda}^K = \sqrt{\frac{\hbar B_{\perp}^2}{4SM\omega_p(\bm{k}\lambda)}}
        \bigg[ik_z\epsilon_{\bm{k}\lambda}^x + k_z\epsilon_{\bm{k}\lambda}^{y} + (ik_x+k_y)\epsilon_{\bm{k}\lambda}^z\bigg]
    \text{ ,}
\end{equation}
we can see that the main difference between the coupling introduced in this work and the phenomenological one (see equations \ref{lambdamec} and \ref{gammamec}), is that the latter comes from an intrinsic mechanism parametrized by the magnetoelastic parameter $B_{\perp}$ and it directly reflects a non-manipulative feature of a particular material. Ultimately, this implies that there is not possibility of manipulating the magnon polarons features as it occurs in the case of the induced magnetoelastic coupling proposed here, which even allows a control level to the point of manipulate the strength of the coupling and choosing which phonon polarizations are coupled to the magnon.

\subsection{Numerical calculations}\label{numerical_calculations}
To obtain the magnon polaron bands and properly compare the contribution of both the phenomenological as the magnetic field gradient-induced magnetoelastic coupling to the system, we perform numerical calculations by employing the Colpa's \cite{colpa1978} algorithm to para-diagonalize Hamiltonian \ref{H_total}.  To implement the algorithm we need to write the Hamiltonian in its quadratic form as
\begin{equation}
    \mathcal{H} = \frac{1}{2}\sum_{\bm{k}}\left[\alpha_{\bm{k}}\dagg\hspace{0.2cm} \alpha_{-\bm{k}}\right]H_{\bm{k}}\left[\alpha_{\bm{k}}\hspace{0.2cm} \alpha_{-\bm{k}}\dagg\right]^T,
\end{equation}
where $\alpha_{\bm{k}} \equiv \big(a_{\bm{k}} ~ c_{\bm{k}1} ~ c_{\bm{k}2} ~ c_{\bm{k}3}\big)$ and $H_{\bm{k}}$ is an $8 \times 8$ hermitian matrix.  Colpa's algorithm will return us a para-unitary matrix $\mathcal{T}_{\bm{k}}$ that satisfies
\begin{equation}
    \label{paradiagonalization}
    \mathcal{T}\dagg_{\bm{k}} H_k \mathcal{T}_{\bm{k}} = \begin{pmatrix}
        E_{\bm{k}} &           0  \\
                 0 & E_{-\bm{k}}
    \end{pmatrix}
    \text{ ,}
\end{equation}
where $E_{\bm{k}}$ is a $4 \times 4$ diagonal matrix containing the eigenenergies. The respective eigenvectors are given by
\begin{equation}
    \begin{pmatrix}
        \bm{\gamma}_{\bm{k}} \\ \bm{\gamma}\dagg_{-\bm{k}}
    \end{pmatrix} = \mathcal{T}_{\bm{k}} \begin{pmatrix}
        \bm{\alpha}_{\bm{k}} \\ \bm{\alpha}\dagg_{-\bm{k}}
    \end{pmatrix}
    \text{ .}
\end{equation}

\section{Magnon polaron bands in Magnonic Crystals}
\label{MPbands}

Here we numerically compute the magnon polaron bands in a Magnonic Crystal embedded in a ferromagnetic insulator. Specifically, we use a YIG sample whose relevant parameters are listed in the Table \ref{parameters_values} \cite{streib2019,flebus2017,ruckriegel2014}. 
\begin{table}[H]
    \centering
    \begin{tabular}{|>{\centering\arraybackslash}m{0.4\columnwidth}|>{\centering\arraybackslash}m{0.4\columnwidth}|@{}m{0pt}@{}}
        \hline
        Parameter     & Value                   &\\[3pt]  \hline
        $S$           & $20$                    &\\[3pt]  \hline
        $M$           & $9.8\times 10^{-24}$kg  &\\[3pt]  \hline
        $a_0$         & $12.376 \angstrom$      &\\[3pt]  \hline
        $v_\parallel$ & $7209$m/s               &\\[3pt]  \hline
        $v_\perp$     & $3843$m/s               &\\[3pt]  \hline
        $J$           & $0.24$meV               &\\[3pt]  \hline
    \end{tabular}
    \caption{Values used in the numerical calculations.}
    \label{parameters_values}
\end{table}
As previously reported, periodicity on a magnetic system gives rise to the so-called Magnonic Crystals \cite{chumak2017}. A Magnonic Crystal can be manufactured by means of periodic modulation on the magnetic anisotropy or magnetic fields \cite{troncoso2015}, periodic inclusion of non-magnetic materials \cite{centa2019}, periodic arrays of dots \cite{gubbiotti2012,tacchi2010} or antidots \cite{ulrichs2010}. Geometrical modulations on the surface of a ferromagnetic film also gives rise to a Magnonic crystal structure \cite{gallardo2018}. The main feature of the Magnonic crystals is the generation of band gaps where spin waves can not propagate, allowing their manipulation for potential devices in spintronic or magnonics. Thus, a Magnonic Crystal can be summarized as a meta-material that enables the suppression and/or propagation of spin waves according to its band structure. Phonons in a crystal also have a band structure, so one would expect that the hybridization between them to be magnified in terms of increasing number of anti-crossing points due to the bands folding. In fact, magnon polarons mediated by the phenomenological magnetoelastic coupling have been recently studied in similar structures \cite{graczyk2017,graczyk20172}.
\newpar

Here we use the arguments presented in Section \ref{model} about the stability of the system (see also the discussion at Appendix \ref{appendix_equilibrium}), which can be summarized as the absence of magnetic field gradient induced-magnon polaron excitations when the applied magnetic field has the same periodicity of the lattice, to explore the magnon polaron excitations in Magnonic Crystals. Recall that the main idea behind using Magnonic Crystals to explore the generation of magnon polarons mediated by a  magnetic field gradients is the fact that we can modulate the magnetic field such that the force exerted by the magnetic field gradient on each unit cell belonging to the Magnonic Crystal is zero. In this way, and as proof of concept, in our numerical calculations, we will use a magnetic field which is likely not easy to experimentally to achieve but that allows us to show how our proposal should work. A more realistic shape of the magnetic field is not a crucial issue in the present formalism because the main importance to have in mind regarding the magnetic field is that it must be in such a way that its gradient cancels the net force on each unit cell whatever the shape it has.
\newpar

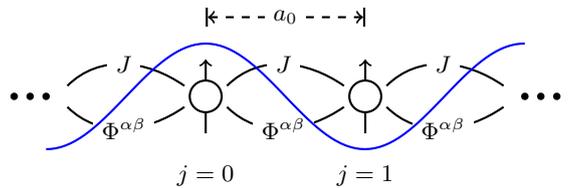
\begin{figure}
    \centering
    \begin{tikzpicture}[thick, scale=0.7]
        \draw[fill] (-0.6,0) circle(0.05);
        \draw[fill] (-0.3,0) circle(0.05);
        \draw[fill] (0.0,0) circle(0.05);
        
        \draw (0.5-0.1,0.2) to[out=45,in=145] node[fill=white,midway] {$J$} (2.5+0.1,0.2);
        \draw (0.5-0.1,-0.2) to[out=-45,in=-145] node[fill=white,midway] {$\Phi^{\alpha\beta}$} (2.5+0.1,-0.2);
        \node at (3,-1.5) {$j=0$};  \draw (3,0) circle(0.3);  \draw (3,-0.7) -- (3,-0.3);  \draw[->] (3,0.3) -- (3,0.7);
        
        \draw[|<->|, dashed] (3,1.5) -- node[fill=white,midway] {$a_0$} (6,1.5);
        
        \draw (3.5-0.1,0.2) to[out=45,in=145] node[fill=white,midway] {$J$} (5.5+0.1,0.2);
        \draw (3.5-0.1,-0.2) to[out=-45,in=-145] node[fill=white,midway] {$\Phi^{\alpha\beta}$} (5.5+0.1,-0.2);
        \node at (6,-1.5) {$j=1$};  \draw (6,0) circle(0.3);  \draw (6,-0.7) -- (6,-0.3);  \draw[->] (6,0.3) -- (6,0.7);
        \draw (6.5-0.1,0.2) to[out=45,in=145] node[fill=white,midway] {$J$} (8.5+0.1,0.2);
        \draw (6.5-0.1,-0.2) to[out=-45,in=-145] node[fill=white,midway] {$\Phi^{\alpha\beta}$} (8.5+0.1,-0.2);

        \draw[fill] (9,0) circle(0.05);
        \draw[fill] (9.3,0) circle(0.05);
        \draw[fill] (9.6,0) circle(0.05);
        
        \draw[domain=0:9,samples=100,blue] plot(\x,{cos(1.05*(\x - 3) r)});
    \end{tikzpicture}
    \caption{Proof of concept of the effect that the particular magnetic field gradient (see Eq. \ref{Bperiodic}) exerts on each site of the unit cell. It can be seen that the spin located at $j=0$ is under an opposite force than the spin located at $j=1$, which ultimately cancels the net force in the unit cell. This is the essence behind the feasibility of our proposal.}
    \label{fig:diagram}
\end{figure}

Thus, in order to adjust the magnetic field to avoid the acceleration on the system, we will use the follow shape of it
\begin{align}
    \label{Bperiodic}
    \bm{B}(x,z) &= \left[\frac{{B'}^{yx}}{q_m}\sin\left(q_m x\right) + {B'}^{yz}z\cos\left(q_m x\right)\right]\bm{\hat{y}}
        \nonumber \\ &
        + B^z \bm{\hat{z}}
    \text{ ,}
\end{align}
where $q_m = 2\pi/(ma_0)$ and $m$ is the number of sites of the basis (see also the related case for $m=1$ depicted in Fig. \ref{Appendix_fig1}). Note that ${B'}^{y\beta}$ ($\beta = x,z$) corresponds to the derivative of the $y$-component of the magnetic field respect to the variable $\beta$, according to our notation prescribed in Eq. \ref{B_notation}. For this particular choice of the magnetic field, we can ensure that a magnetic field gradient will not produce a net force on the unit cell as long as $m$ is an even number.  Fig. \ref{fig:diagram} shows, with a solid blue line the $x$-component (modulus) of the magnetic field gradient, as derived from Eq. \ref{Bperiodic}, as a function of the distance along the $x$ direction. According to our proposal, from Fig \ref{fig:diagram} it can be seen that the exerted force driven by the magnetic field gradient on the spin located at $j=0$, points in the opposite direction than the exerted one on the spin located at $j=1$, canceling thus the net force on the unit cell.

\begin{figure*}
\centering
\includegraphics[width=0.8\textwidth]{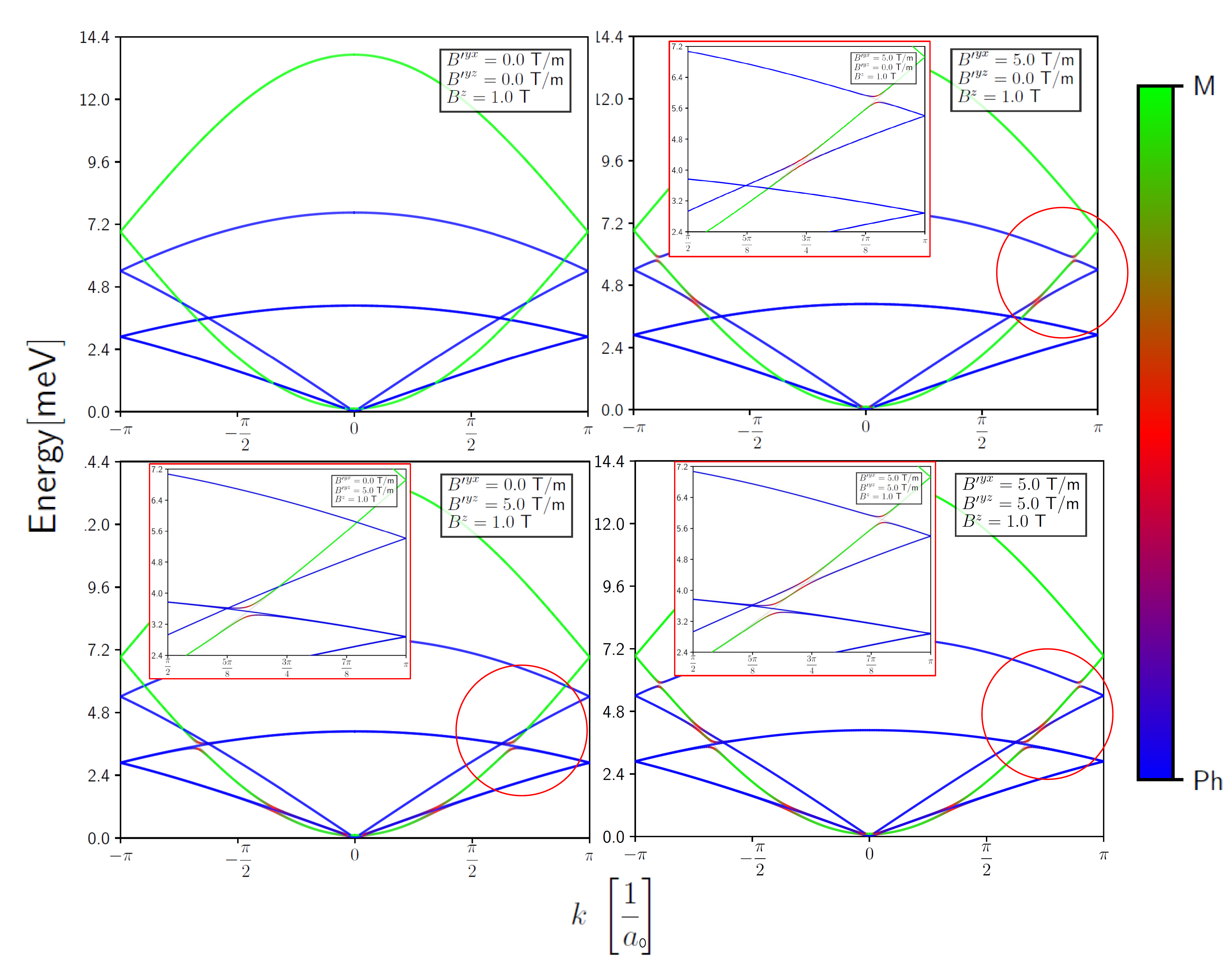}%
\caption{Magnon polaron bands for a YIG Magnonic Crystal with wave vector $\bm{k}\parallel\bm{\hat{x}}$ and a magnetic field $\bm{B}(x,y)=(0,B^y(x,z),1)$ T as mentioned in the main text. The number of the sites in each unit cell is set $m=2$. The color bar shows the amplitude of the probability of the magnon polaron wave function such that the green line corresponds to a quasi full magnon band, while the blue line corresponds to a quasi full phonon one. The maximum mixture between magnons and phonons is represented in red color.  a) Magnon polaron bands for ${B'}^{yx} = {B'}^{yz} = 0$, b) for ${B'}^{yx} = 5$ T/m and ${B'}^{yz}=0$, c) for ${B'}^{yx}=0$ and ${B'}^{yz} = 5$ T/m; and d) ${B'}^{yx} = {B'}^{yz} = 5 \text{ T/m}$. The wave vector $\bm{k} = k\bm{\hat{x}}$ is written in units of $1/a_0$. The insets in the red boxes are a zoom-in of the anti-crossing points marked with a red circle.}
\label{Fig3}
\end{figure*}

Fig. \ref{Fig3} shows the magnon polaron bands for waves propagating along the $\bm{\hat{x}}$-direction in absence of the phenomenological magnetoelastic coupling in a Magnonic Crystal embedded in a YIG sample with $m=2$, for different values of the magnetic field gradient ${B'}^{y\beta}$ presented in Eq. \ref{Bperiodic}, and a constant magnetic field of $B^z = 1$ T applied into the $\bm{\hat{z}}$-direction.  The color bar is a representation of the amplitude of the probability of which character has the wave function and its corresponding eigenenergy. In this way, the green color represents essentially a magnon state, while the blue color a phonon state. The intermediate colors show how mixed are magnons and phonons, reaching the maximum coupling at the red color when $\bm{k} = \bm{k}_i^{*}$, being $\bm{k}_i^{*}$ the $ith$ anti-crossing point. Note that the whole spectrum corresponds to magnon polaron excitations, but far from the anti-crossing point these excitations behave like non-interacting magnons or phonons.
\newpar

Figure $\ref{Fig3}$a) shows the magnon and phonon dispersions in the first Brillouin zone with no magnetoelastic coupling since the magnetic field gradient turns to be zero (${B'}^{y\alpha} = 0$). As expected, the magnon and phonon bands do cross, which reflects the absence of interaction between them. Next, we turn on the magnetic field gradient along the longitudinal direction, as depicted in figure \ref{Fig3}b), with ${B'}^{yx} = 5 $ T/m and ${B'}^{yz} = 0$. The appearance of distinct band gaps at the crossing points is evident (so anti-crossing points), which manifests the coupling between the longitudinal phonon mode with magnons propagating along the $\bm{\hat{x}}$ direction. Note that none of the transverse phonon modes couple with magnons as pointed out in Eq. \ref{lambdamec}. Also, it can be seen that due to the band folding effect, acoustic magnons might simultaneously couple with acoustic and optical phonons. Analogously, Fig \ref{Fig3}c) shows the magnon polaron bands for a magnetic field gradient applied into the $\bm{\hat{z}}$-direction with ${B'}^{yz} = 5$ T/m and ${B'}^{yx} = 0$. Similarly to the previous case, here magnons only couple with a transverse phonon mode and again simultaneously couple with acoustic and optical phonons. Note that the energy band gaps $\Delta_i$ between the magnon polaron modes at $\bm{k}^{*}_i$ are different for the cases with the gradient applied into the $\bm{\hat{x}}$ and $\bm{\hat{z}}$ directions, as will be shown in detail below.
% \newpar

\newpar
Furthermore, and in order to depict the ability to control which magnon and phonon bands hybridize, Fig. \ref{Fig3}d) shows the system under the action of a magnetic field gradient applied into the transverse and longitudinal directions with ${B'}^{yz} = {B'}^{yx} = 5$ T/m. In this case, there is a coupling of both distinct phonon modes with magnons propagating along the $\bm{\hat{x}}$ direction since the magnetic field is applied into two distinct spatial directions. If we would have considered a magnetic field gradient applied into the three spatial directions, then the degenerated transverse phonon mode of YIG should be also coupled and, consequently, obtaining two degenerated magnon polaron bands. In the same way, for a given material with three distinct phonon modes, a magnetic field gradient with the three spatial components gives rise to three different magnon polaron bands. Then, we have shown that it is possible to choose which magnon polaron mode to excite by only fixing the direction of the magnetic field gradient. Remarkably, under the same conditions as above, but considering only the phenomenological magnetoelastic coupling, the ability to choose which magnon and phonon bands interact does not exist.
\newpar

\begin{figure}
    \centering
    \includegraphics[width=0.48\textwidth]{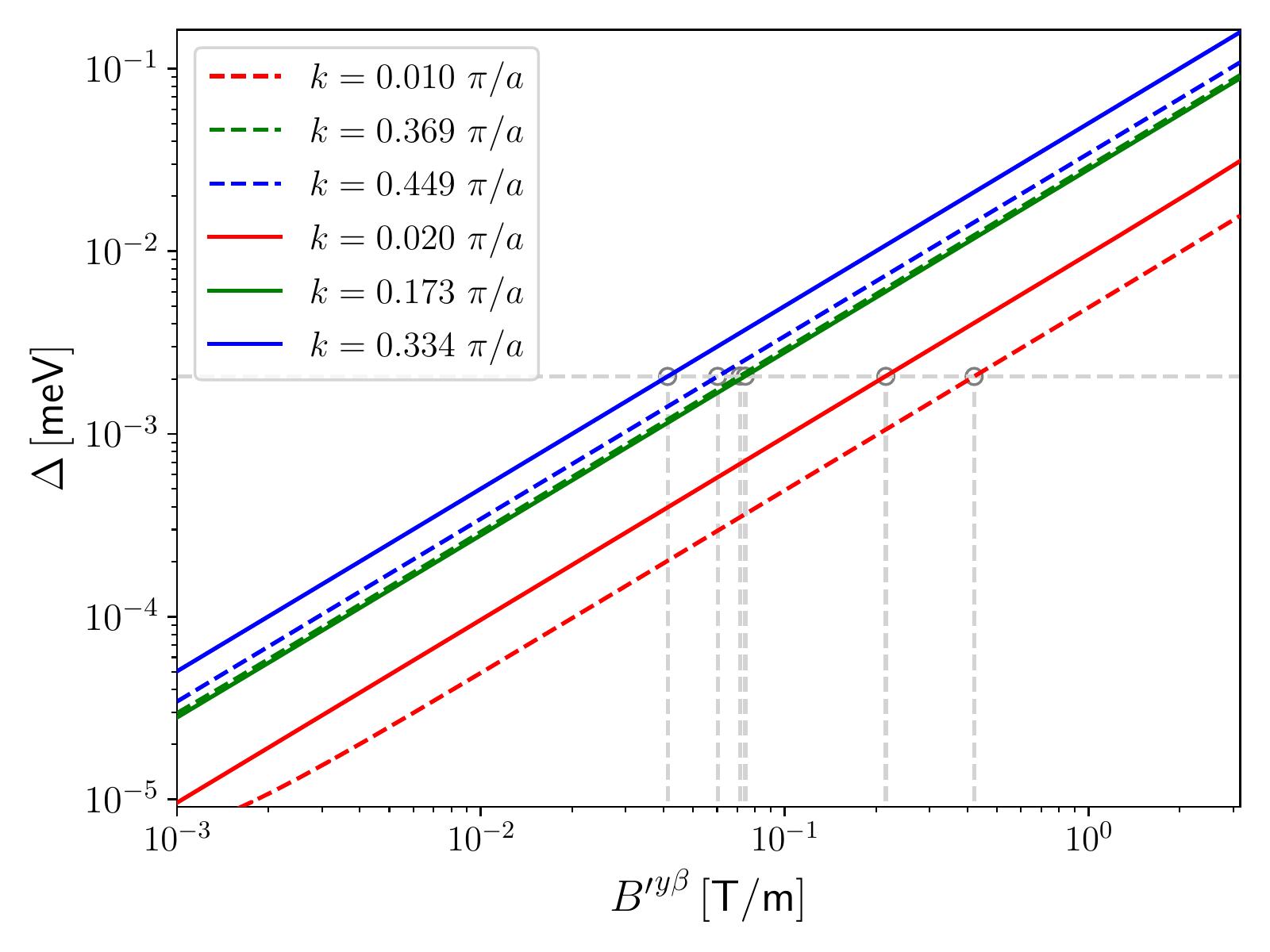}%
    \caption{Energy gap for different values of the magnetic field gradints ${B'}^{y\beta}$ presented in equation \ref{Bperiodic} in the main text for a YIG Magnonic Crystal with a basis $m=2$. In solid and dashed lines it is shown the gap between acoustic magnons and transverse and longitudinal phonon modes, respectively. In this specific case and due to the band folding effect, the blue lines represent coupling between acoustic magnons and optical phonons, while red and green lines are used to show the couplingb between acoustic magnons and phonon modes. The grey horizontal line represents the energy gap $\Delta_K$ obtained in the reference \cite{flebus2017}.}
    \label{Fig4}
\end{figure}

To compare the contribution of this induced magnetoelastic coupling with the usual phenomenological one, we compute the band gap $\Delta_i = E^{\text{MP}_1}_{\bm{k}}-E^{\text{MP}_2}_{\bm{k}}\vert_{\bm{k} = \bm{k}^{*}_i}$ that separately generates each \textit{kind} of coupling between the magnon polaron bands at the $ith$ anti-crossing point $\bm{k}^{*}_i$ and using the same parameters as above, i.e., a YIG Magnonic Crystal with $m=2$. Thus, in Fig. \ref{Fig4}, we have plotted in a log-log scale the energy gap $\Delta$ for different values of the magnetic field gradient ${B'}^{y\beta}$ ($\beta = x,z$). As stated above, the gap is measured at all the possible wave vectors $\bm{k} = \bm{k}^{*}$ where the magnon and phonon bands would cross in the absence of magnetoelastic coupling and have been depicted by solid and dashed lines according the coupled phonon mode. Specifically, the solid lines represents coupling between magnons and transverse phonon modes, whereas the dashed lines corresponds to coupling between magnons and longitudinal phonon modes. The red, blue, and green colors are used to show in which anti-crossing point $\bm{k}^{*}_i$ is measured the energy band gap. Importantly, in this case, the largest anti-crossing points are a consequence of the band folding effect meaning that optical phonons couple with acoustic magnons (see Fig. \ref{Fig3}) and it can be seen that they generate the largest band gaps too. In this plot, the log-log scale has been used to show the energy gap for a wide range of magnetic field gradients, however, due to the evident linear behavior of $\Delta_i$, we have also performed a linear plot depicted at Appendix \ref{Delta_gradient} in Fig. \ref{Appendix_fig2}. It is of particular interest to find the value of ${B'}^{y\beta}$ for which each gap reproduces the gap seen in \cite{flebus2017}, which is approximately $\Delta_K\approx 2.1\mu$eV, an aspect that has been marked by the grey dotted lines. The importance of reproducing such a gap by means of the current proposal relies on the fact that most of the experimental measurements have been accomplished in YIG samples by only considering the Phenomenological contribution. 
\newpar

From Fig. \ref{Fig4} we can also see with solid lines that acoustic magnons couple with longitudinal acoustic phonons at $k^{*}_{\parallel 1}\approx 2.538\times 10^{7}$ m$^{-1}$ and $k^{*}_{\parallel 2}\approx 9.367\times 10^{8}$ m$^{-1}$; whereas acoustic magnons couple transverse acoustic phonons at $k^{*}_{\perp 1}\approx 5.077\times 10^{7}$ m$^{-1}$ and $k^{*}_{\perp 2}\approx 4.392\times 10^{8}$ m$^{-1}$. Similarly, Fig. \ref{Fig4} also shows that at $k^{*}_{\parallel 3}\approx 1.134\times 10^{9}$ m$^{-1}$ acoustic magnons and a longitudinal optical phonon mode are coupled, while at $k^{*}_{\perp 3}\approx 8.478\times 10^{8}$ m$^{-1}$ acoustic magnons couple with the transverse optical phonon band. Importantly, the gap $\Delta_K$ can be reproduced with magnetic field gradients of about $\sim 0.1$ T/m, that in fact can be reasonably achieved with the current experimental techniques \cite{kunert2014,johanning2009,welzel2011,khromova2012,mintert2001}. Indeed, by using a magnetic microtrap, the reference \cite{reichel1999} reported gradients of up to $8000$ T/m, which due to the linear behaviour of $\Delta$ with ${B'}^{\alpha\beta}$, it would translates into very large $\Delta$ values, so we predict enhancements of transport properties related to the presence of magnon polarons when a magnetic field gradient larger than $0.1$ T is applied. Interestingly, Fig. \ref{Fig4} (see also Fig. \ref{Appendix_fig2}) also shows that we could reach gaps of order of about $0.1$ meV with gradients of about $\sim 10 \text{ T/m}$. This gap is of the same order as reported on previous works on topological magnonics \cite{diaz2019}.
\section{Conclusions}
\label{discussion}
In this work, we have proposed a versatile way to induce a magnetoelastic coupling in any (anti)ferromagnetic material. Despite our formalism was developed in a specific spin chain, it can be extended to more sophisticated systems where we expect similar behaviors due to the generalized treatment we gave for the total Hamiltonian. The main contribution of the present work is the proposal of an enhancement of the magnetoelastic coupling by a magnetic field gradient. The physics behind it can be understood in terms of the force exerted by the magnetic field gradient on each magnetic dipole which deviates them from its equilibrium position exciting thus simultaneously both phonon and magnon modes. Importantly, the order of magnitude of the magnetic field gradient needed to achieve measurable effects starts from $\sim 10^{-1}$ T/m in YIG, which is very well accomplished in standard experiments. Since a magnetic field gradient means an external force on each magnetic dipole, an infinite system with the same periodicity of the magnetic field is then accelerated and a non-hermitian Hamiltonian is expected when considering the $\hat{\bm{z}}$-axis as ground state, so no magnon polarons can be excited. This can be overcome by properly adjusting the magnetic field periodicity such that the net force on each unit cell of the system is zero. Thus, by employing our proposal in a Magnonic Crystal, where the nature of it allows having such features, we can avoid the imaginary parts of the energy spectrum and real energies are obtained in the whole Brillouin zone. Furthermore, as a highlihgted results, the band gaps in the Magnonic Crystal can be controlled by varying the strength and direction of the magnetic field gradient. Note that since the formalism demands a stable ground state in the system, which must be accomplished canceling the net force emerged from the gradient on the unit cell of the system, our proposal should be very well achieved in an antiferromagnetic system, where the nature of the unit cell would allow major liberty on the choice of the magnetic field shape.
Finally, our proposal could open new possibilities to control the magnon-phonon interaction with the idea of manufacturing efficient spintronic ans/or magnonics devices. We claim then that this induced magnetoelastic coupling is fully controllable by a magnetic field gradient. Since most of magnon polaron transport properties depend on the strength of this interaction, we predict thus an enhancement of them by controlling the strength and direction of the magnetic field gradient.

\section{Acknowledgments}
N. V-S thanks Fondecyt Postdoctorado Nº 3190264. ASN thanks Fondecyt Regular Nº 1190324. This project has received funding from the European Research Council (ERC) under the European Union’s Horizon 2020 research and innovation programme (grant agreement No. 725509).

\newpar
N. Vidal-Silva and E. Aguilera contributed equally to this work.

\counterwithin{figure}{section}
\appendix

\section{Equilibrium condition in an arbitrary lattice}
\label{appendix_equilibrium}

To analyze the equilibria in an arbitrary lattice, let us firstly start by analyzing the equilibrium in a system composed of a single spin attached to a spring and coupled to a inhomogeneous magnetic, whose Hamiltonian is described by:
\begin{equation}
    \label{eq:single_spinH}
    \mathcal{H} = \frac{\bm{p}^2}{2M} + \frac{M {\omega_0}^2}{2} \bm{u}^2 - \mu_B g \bm{B} \cdot \bm{S}
    \text{,}
\end{equation}

where we have defined the small deviation $\bm{u} = (x - X_0)\bm{\hat{x}}$, being $X_0$ the equilibrium position. Classically, the spin's time evolution is governed by Newton's second law and Landau-Lifschitz-Gilbert equation. To make the study of system's equilibria more comfortable, the spin variable can be written in spherical coordinates angles $\theta, \phi$ and, it can be considered the particular case where $\bm{B} = (B^x(x), 0, B^z)$, meaning that the classical energy $E(x,\theta,\phi)$ is given by:
\begin{eqnarray}
    \label{eq:single_spinH_1}
  \nonumber E(x,\theta,\phi) = \frac{\bm{p}^2}{2M} + \frac{M {\omega_0}^2}{2} \bm{u}^2\\
  -\mu_B g S B^x \sin(\theta)\cos(\phi) - \mu_B g S B^z \cos(\theta)
    \text{ .}
\end{eqnarray}

From equation~\ref{eq:single_spinH_1}, the equilibria of the system follow directly from minimizing the energy. For this, it must be imposed that $\partial_x E = 0$, $\partial_\theta E = 0$ and $\partial_\phi E = 0$:
\begin{subequations}
    \label{eq:spin_equilibrium}
    \begin{align}
        M {\omega_0}^2 u - \mu_B g S \frac{\partial B^x}{\partial x} \sin(\theta)\cos(\phi) &= 0  \\
        B^x \cos(\theta) \cos(\phi) - B^z \sin(\theta) &= 0  \\
        B^x \sin(\theta)\sin(\phi) &= 0
        \text{.}
    \end{align}
\end{subequations}

From equation~\ref{eq:spin_equilibrium}, it is direct to see that $(u, \theta, \phi) = (0, 0, 0)$ is a solution.  Studying the Hessian at the equilibrium point, it is clear that this is positive definite whenever the following inequality complies:
\begin{equation}
    \label{eq:spin_stable_equilibrium}
    \Bigg( \frac{\partial B^x}{\partial x}\bigg|_{u=0} \Bigg)^2 < \frac{M \omega_0^2}{g \mu_B S} B^z
\end{equation}

Thus, the point $(u, \theta, \phi) = (0, 0, 0)$ is a stable equilibrium whenever the condition established by inequality~\ref{eq:spin_stable_equilibrium} is fulfilled.  The importance of analyzing the equilibrium of Hamiltonian~\ref{eq:single_spinH} comes from the fact that the equilibrium needs to be stable for it to have spin waves.  This single spin-toy model allows us to see that there exists a magnetic field gradient from which the equilibrium is no longer stable and no spin waves are admitted under those conditions.
\newpar

Now we go beyond this single spin model and expand our analysis of equilibrium to an extended one-dimensional lattice with spacing $a_0$ parameter. The simplest system with both magnetic and elastic interaction is given by:
\begin{eqnarray}
    \label{eq:spin_lattice_H}
    \nonumber\mathcal{H} = \sum_i\Bigg[
        \frac{{\bm{p}_i}^2}{2M} + \frac{M {\omega_0}^2}{2} (\bm{u}_{i+1}-\bm{u}_i)^2 - J \bm{S}_i \cdot \bm{S}_{i+1}\\
        - \mu_B g \bm{B} \cdot \bm{S}_i
    \Bigg]
    \text{,}\hspace{0.2cm}
\end{eqnarray}
\noindent
where the unidimensional elastic interaction and Heisenberg exchange were considered.  To further simplify the model, the magnetic field $\bm{B}(x)$ will be assumed to be periodic with a periodicity equal to that of the lattice. Next, we expand the arbitrary magnetic field $\bm{B}(x)$ around the equilibrium positions $\bm{X_i}$ up to first order in the displacement $\bm{u}$, which leads to:
\begin{eqnarray}
    \label{eq:spin_lattice_expansionH}
    \nonumber\mathcal{H} = \sum_i\Bigg[
        \frac{{\bm{p}_i}^2}{2M} + \frac{M {\omega_0}^2}{2} (\bm{u}_{i+1}-\bm{u}_i)^2 - J \bm{S}_i \cdot \bm{S}_{i+1}\\
        - \mu_B g \bm{B} \cdot \bm{S}_i - \mu_B g\frac{\partial\bm{B}}{\partial x^\mu} \cdot \bm{S}_i u^\mu
    \Bigg]
    \text{ ,}\hspace{0.2cm}
\end{eqnarray}
\noindent
where the positions $\bm{X}_{i} = (ia_0, 0, 0)$ and the spin ground state $\bm{S}_{i} = (0, 0, S)$ were chosen as possible equilibria of the system.
\newpar

To study the actual equilibria of Hamiltonian~\ref{eq:spin_lattice_H}, the spin variable will be written in terms of its spherical angles $\theta_i$ and $\phi_i$ and, as in the single spin case, it will be assumed that the magnetic field is given by $\bm B = (B^x(x), 0, B^z)$.  With these considerations, the Hamiltonian can be written explicitly as:
\begin{eqnarray}
    \label{eq:spin_lattice_explicitH}
    \nonumber\mathcal{H} = \sum_i \Bigg[\frac{{\bm{p}_i}^2}{2M} + \frac{M {\omega_0}^2}{2} (\bm{u}_{i+1}-\bm{u}_i)^2 \\
    \nonumber-JS^2 \bigg(\sin(\theta_i)\sin(\theta_{i+1})\cos(\phi_i)\cos(\phi_{i+1})\\
    \nonumber+ \sin(\theta_i)\sin(\theta_{i+1})\sin(\phi_i)\sin(\phi_{i+1})+\cos(\theta_i)\cos(\theta_{i+1})\bigg)  \\ 
    - \mu_B g \bigg( SB^x(x)\sin(\theta_i)\cos(\phi_i) + SB^z\cos(\theta_i) \bigg)
    \Bigg]\text{.}\hspace{0.6cm}
\end{eqnarray}

Using the single spin case as inspiration, the point $(u_i, \theta_i, \phi_i) = (0, 0, 0)$ is considered as possible equilibrium.  We expand the Hamiltonian~\ref{eq:spin_lattice_explicitH} to second order around the mentioned point:
\begin{eqnarray}
    \label{eq:spin_lattice_taylorH}
    \nonumber\mathcal{H} = \sum_i \Bigg[
        \frac{{\bm{p}_i}^2}{2M} + \frac{M {\omega_0}^2}{2} (\bm{u}_{i+1}-\bm{u}_i)^2\\
        - JS^2 \Big(\theta_i\theta_{i+1} - {\theta_i}^2 - {\theta_{i+1}}^2\Big) - \mu_B g \bm{B} \cdot \bm{S}_i\Bigg]\text{.}
\end{eqnarray}

\begin{figure} 
\centering
\includegraphics[width=0.48\textwidth]{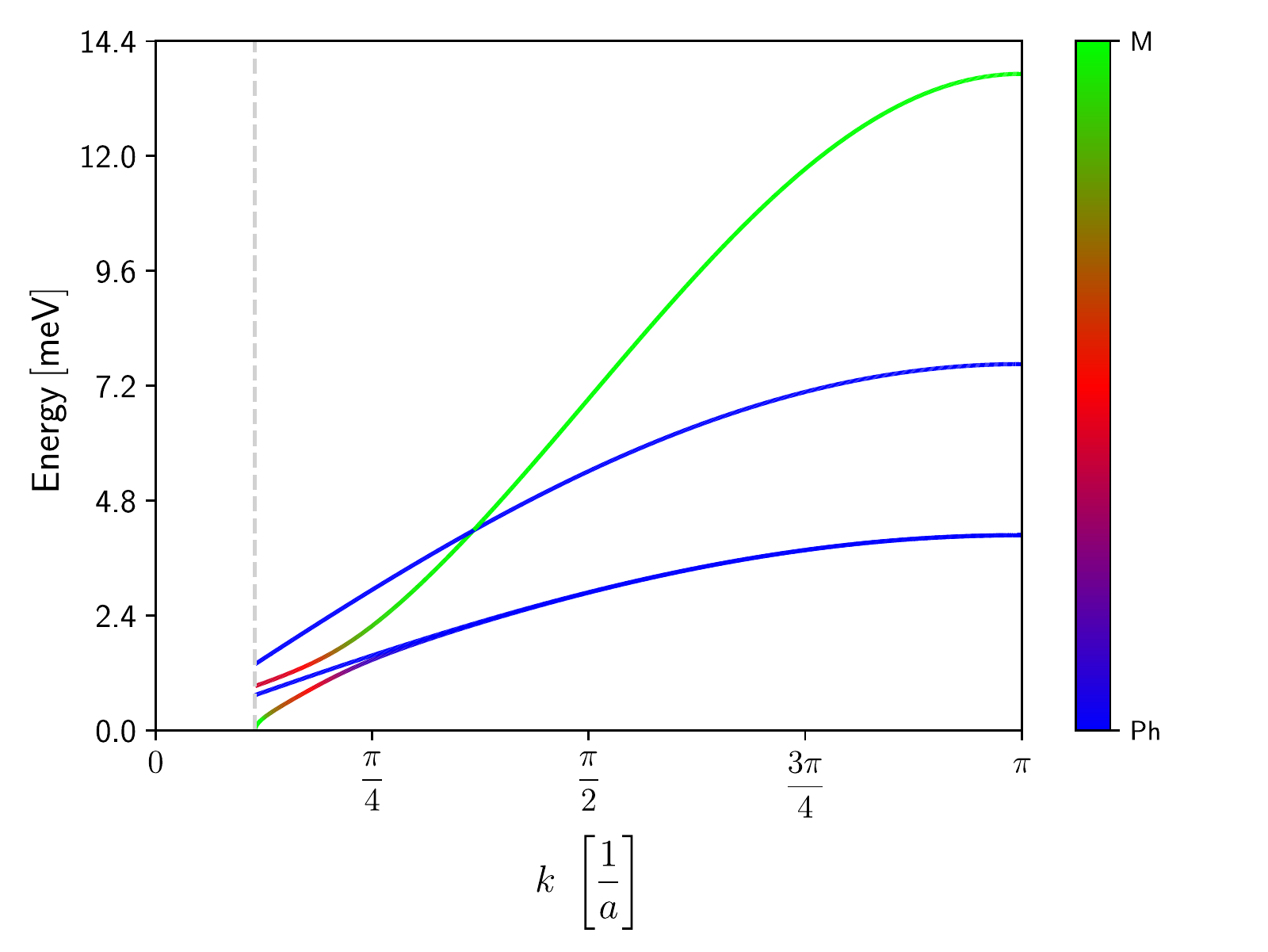}
\caption{Magnon polaron bands for a YIG sample as function of $1/a_0$ for $B^x = 1$ nT. The green color is used to denote the
probability amplitude of the eigenstate to by a magnon and blue is used to denote any phonon mode. The grey-dotted vertical line denotes the point when the Hamiltonian is no longer positive definite and no physical eigenenergies are observed below it.}
\label{Appendix_fig1}
\end{figure}

To find the stability of the system, Fourier's theorem on $u_i$ and $\theta_i$ will be used.  This will lead to a Hamiltonian described by $u_k$, $u_{-k}$, $\theta_{k}$ and $\theta_{-k}$, which is given by:
\begin{eqnarray}\label{eq:spin_lattice_blochH}
\nonumber\mathcal{H} = \sum_k \Bigg[\frac{{p_k}^2}{2M} + M {\omega_0}^2 \Big(1 - \cos(ka)\Big) u_k u_{-k} \\
\nonumber- 2JS^2 \Big(1 - \cos(ka)\Big) \theta_k \theta_{-k}+ \mu_B g S B^z \theta_k \theta_{-k}\\ - \mu_B g S \frac{\partial B^x}{\partial x} u_k \theta_{-k}\Bigg]
    \text{ .}
\end{eqnarray}
From the Hamiltonian in k-space, it can be noted that the Hessian is a block diagonal matrix, where each block is $4\times4$, formed by the variables $u_k$, $\theta_k$, $u_{-k}$ and $\theta_{-k}$.  With this, it can be proved that every eigenvalue is positive if and only if
\begin{eqnarray}
    \label{eq:lattice_equilibrium}
    \nonumber\bigg(B'Sg\mu_B\bigg)^2 < 4mS{\omega_0}^2 \Big(-2JS\big(1 - \cos(a k)\big)\\
    + \mu_B g B^z\Big) \sin^2\bigg(\frac{a k}{2}\bigg)\text{ ,}\hspace{0.5cm}
\end{eqnarray}
which is Eq. \ref{equilibrium_condition} of the main text. As an extra note, it is essential to recognize that, intuitively, the proposed equilibrium could never be stable because every site feels the same force, thus always pushing it away from the equilibrium obtained with a constant magnetic field. As explained in section \ref{numerical_calculations} of the main text, the way we choose to diagonalize the Hamiltonian of the system is following the Colpa's algorithm, which demands to write the total Hamiltonian into its quadratic form. However, as shown in Eq. \ref{eq:lattice_equilibrium}, there will be some range of $k$ values where the diagonalization procedure breaks down essentially because of the system is no longer stable at the chosen equilibrium points. In order to show explicitly such a claim, in Fig. \ref{Appendix_fig1} we compute the magnon polaron bands for a magnetic field given by
\begin{eqnarray}
    \bm{B}(x,y) = B^x\sin\left(\frac{2\pi}{a_0}x\right)\bm{\hat{x}}
    \text{ ,}
\end{eqnarray}
where must be noticed that $\bm{B}(x,y)$ has the same periodicity of the lattice, in the same way as assumed to arrive to Eq. \ref{eq:lattice_equilibrium}.
Also, the magnetic field gradient has been applied into the $x$-direction, it has been used $B^x = 1$ nT, and the magnetic parameters were extracted from the Table \ref{parameters_values} for a YIG sample. Importantly, the grey-dotted vertical line marks the point where inequality \ref{eq:lattice_equilibrium} stops being true. Thus, it has been proven that when applying a magnetic field gradient on a magnetic field with the same periodicity of the lattice, the system feels a net force which ultimately accelerates it and is no longer stable. This has the consequence of magnon polarons excitations should not be observed for this case.
\section{Energy gap $\Delta$ as a function of the magnetic field gradient}\label{Delta_gradient}
\begin{figure} [h]
\centering
\includegraphics[width=0.48\textwidth]{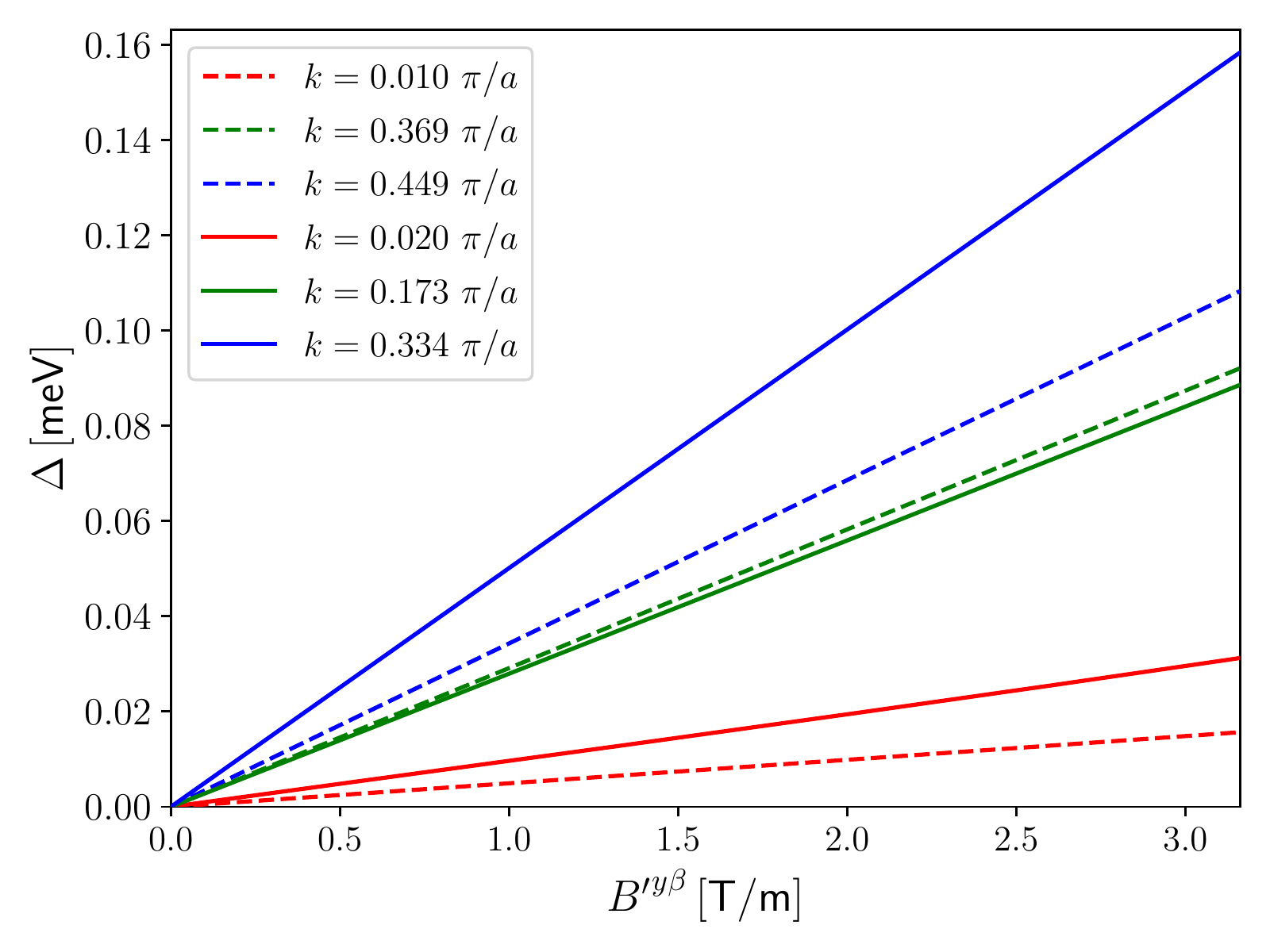}%
\caption{Energy band gap $\Delta$ as a function of the magnetic field gradient ${B'}^{y\alpha}$ as depicted in Fig. \ref{Fig4}. In this case, the linear scale has been used. }
\label{Appendix_fig2}
\end{figure}

Here we show the dependence of the energy gap $\Delta$ as a function of the magnetic field gradient. As depicted in Fig. \ref{Fig4} of the main text, the log-log scale was used to show a wide range of magnetic field gradient. However, here we use a linear scale to show that the energy gap $\Delta$ effectively varies linearly with ${B'}^{y\beta}$, as shown in Fig. \ref{Appendix_fig2}.
The most important aspect of Fig. \ref{Appendix_fig2} is that a linear equation can directly describe it:
$\Delta = b{B'}^{\mu}$ ($\mu=\alpha,\beta$), and the log-log scale is not necessary to show the linear behavior of the energy gap but was nonetheless used to clearly show in the same plot the value upon which the gap in reference \onlinecite{flebus2017} is obtained.
\bibliography{MP}

\end{document}